\documentclass[preprint]{sigplanconf}

\usepackage{amsmath, amssymb, amscd}
\usepackage{amsfonts}
\usepackage{graphicx}
\usepackage{float}
\usepackage{multirow}
\usepackage{times}
\usepackage{listings}
\usepackage{verbatim}
\usepackage{color}
\usepackage{xcolor}
\usepackage{courier}
\usepackage{epsfig}
\usepackage{txfonts}
\usepackage{hyperref}
\usepackage{breakurl}
\usepackage{url}
\usepackage{placeins}
\usepackage{listing}
\usepackage[normalem]{ulem}     

\usepackage[yyyymmdd,hhmmss]{datetime}

\usepackage{oldlfont,epsfig}
\usepackage{epic}
\usepackage{eepic}

\newcommand{\Xomit}[1]{}
\newcommand{\ignore}[1]{}
\newcommand{\Elide}[1]{}

\begin{document}

\setlength{\pdfpageheight}{\paperheight}                                             
\setlength{\pdfpagewidth}{\paperwidth}                                               
                                                                                     
\conferenceinfo{Transact 2015} {June 15--16, 2015, Portland, Oregon, USA.}
\copyrightyear{2015}
\copyrightdata{978-N-NNN-NNNN-N/NN/NN}
\doi{nnnnnn.nnnnnn} 

\title{The Influence of Malloc Placement on TSX Hardware Transactional Memory} 


\authorinfo{Dave Dice}
           {Oracle Labs}
           {dave.dice@oracle.com}
\authorinfo{Tim Harris}
           {Oracle Labs}
           {timothy.l.harris.@oracle.com}
\authorinfo{Alex Kogan}
           {Oracle Labs}
           {alex.kogan@oracle.com}
\authorinfo{Yossi Lev}
           {Oracle Labs}
           {yossi.lev@oracle.com}

\date{\today}
\maketitle




\newcommand\Invisible[1]{
  {\color{white}{\fontsize{1}{1}\selectfont {#1}}}
}

\newcommand\InvisibleVA[1]{
  {\color{white}{\fontsize{1}{1}\selectfont {[#1]}}}
}

\newcommand{\InvisibleVB}[1]{
  {\par {\color{white}\fontsize{1}{1}\selectfont #1 \par}}  
}

\newcommand{\InvisibleVC}[1]{
  {\begin{minipage}{\textwidth} {\color{white}\fontsize{1}{1}\selectfont #1} \end{minipage}}  
}

\newcommand\InvisibleVerbatimB[1]{
  \begingroup 
  \color{white}\fontsize{1}{1}\selectfont 
  \begin{verbatim}
  {#1} 
  \end{verbatim}
  \endgroup
}

\newcommand\InvisibleVerbatimC[1]{
  \begingroup \color{white}\fontsize{1}{1}\selectfont \texttt{#1} \endgroup
}
 
\newcommand\InvisibleVerbatim[1]{
  \begin{minipage}{\textwidth} {\color{white}\fontsize{1}{1}\selectfont \texttt{#1}} \end{minipage} 
}
 
\newcommand\InvisibleQuote[1]{
 \begingroup 
 \color{white}\fontsize{1}{1}\selectfont 
 \begin{quote} {#1} \end{quote} 
 \endgroup
}

\newcommand{\RedText}[1]{\textcolor{red}{[#1]}}


\newcommand{\malloc}{{\footnotesize \texttt{malloc}}}

\begin{abstract}
The hardware transactional memory (HTM) implementation in Intel's i7-4770 
``Haswell'' processor~\cite{Haswell, RitsonBarns} tracks the transactional read-set in 
the L1 (level-1), L2 (level-2) and L3 (level-3) caches \footnote{We have observed read-only
transactions with a cache footprint of 7.5MB successfully commit, but have never
seen successful transactions larger than 8MB -- the size of the L3 cache.} 
and the write-set in the L1 cache.  Displacement or eviction
of read-set entries from the cache hierarchy or write-set entries from the L1 results in abort.  
\footnote{If the line is evicted, the processor loses the ability to track the locations for
conflicts, so it aborts.} 
We show that the placement policies of dynamic storage allocators -- such as those 
found in common \malloc{} implementations -- can
influence the L1 \emph{conflict miss}~\cite{Hill89} rate in the L1~\cite{ISMM11-Afek-CIA}.
Conflict misses -- sometimes called \emph{mapping misses} -- arise because of less 
than ideal associativity and represent imbalanced distribution of active memory
blocks over the set of available L1 indices.  
Under transactional execution conflict misses may manifest as aborts, representing wasted
or futile effort instead of a simple stall as would occur in normal execution mode.   

Furthermore, when HTM is used for \emph{transactional lock elision} (TLE) 
~\cite{asplos09-dice, SMLI-TR-2009-180}, persistent aborts arising from 
conflict misses can force the offending thread through the so-called ``slow path''.  
The slow path is undesirable as the thread must acquire the lock
and run the critical section in normal execution mode, precluding the
concurrent execution of threads in the ``fast path'' that monitor that same lock
and run their critical sections in transactional mode~\cite{RWTLE}. 
For a given lock, multiple threads can concurrently use the transactional fast path,
but at most one thread can use the non-transactional slow path at any given
time.  Threads in the slow path preclude safe concurrent fast path 
~\cite{wttm14-dice} execution.  Aborts rising from placement policies and
L1 index imbalance can thus result in loss of concurrency and reduced aggregate
throughput. 

We demonstrate that allocator placement policies can influence aborts arising
from index conflicts, and that \emph{index-aware} allocators can serve to 
reduce the incidence of such aborts.  

\Invisible{Data accessed in the critical section body} 
\Invisible{We have observed read-only transactions commit where the
cache footprint exceeds the L2 size, but have never observed transactions
commit where the footprint is above the size of the L3}

\begingroup
\color{white}\fontsize{1}{1}\selectfont 
\begin{verbatim}

cat pigdog donkey

testline

\end{verbatim}
\endgroup

\InvisibleQuote{

*   with TSX-RTM, misses turn into aborts 
    aborts : amplify impact of misses; wasted cycles
    aborts also force slow-path and loss of concurrent execution for TLE

*   ringer : expected inequalities
    +   rate = 1 / time
    +   Expect : Inequality-A
        Rate(TTS)/Rate(TLED) >> Rate (TTS+CIA)/Rate(TLED+CIA) 
    +   expect : inequality-B
        Rate(TTS+CIA)/Rate(TTS) << Rate(TLED+CIA)/Rate(TLED)
        Inequality-A is equivalent to inequality-B
    +   CIA benefits
        Benefit : reduced misses
        Benefit : reduces misses and reduced aborts
    +   expect : Rate(TTS+CIA) >> Rate(TTS)
        expect : Rate(TLED+CIA) >> Rate(TLED)
        expect : Rate(TTS+CIA)/Rate(TLED+CIA) near 1.0

    +   Ideally for paper : to support claims 
        Pick configuration to satisfy inequality-B and satisfy the following :
        Rate (TTS) >> Rate(TLED) 
        Rate (TTS+CIA) approx Rate(TLED+CIA)
        Rate (TTS+CIA) >> Rate(TTS) >> Rate(TLED) 

}

\end{abstract}




\section{Introduction}

For background, the Intel i7-4770 processor has a relatively simple L1 cache geometry. 
The L1 data cache is 32KB with 64-byte lines, physically tagged, 
8-way set-associative. There are 64 possibly indices (sets). As such the 
\emph{cache page} size is 4KB -- addresses that differ by an integer multiple 
of 4K will map to the same index (set) in the L1 and compete for the 8 lines within that set.  
The L1 contains 512 lines.  
Each core has private L1 and L2 caches, while the L3 is shared by all cores
on the chip.  The L2 and L3 are unified -- able to contain both code and data. 
The L2 instances are 256KB each and 8-way set-associative, and the single common 
per-chip L3 is 8MB and also 8-way set-associative \footnote{We were unable to determine 
inclusivity relationships between the L1, L2 and L3}.  
The low-order 6 bits of the 
address presented to the L1 form the offset into the line, and the next higher 
6 bits serve as the L1 index. The MMU base page size is 4KB, so there is no 
overlap between the virtual page number and the L1 index field in a virtual address. 
The L1 index field passes through address translation verbatim \footnote{
We assume the x86 segment descriptor base addresses are set to 0}. As such, 
operating system-level page coloring~\cite{Romer94} is not effective in the L1. 
(An advantage of this design is that indexing can commence before the virtual 
address is translated to a physical address, although the cache still ultimately needs the physical 
address for tag comparison). Some CPUs hash addresses~\cite{Hund} -- usually XORing 
high-order physical address bits into the index bits -- in order to reduce the 
odds of index hotspots and imbalance, but experiments suggest that does 
not appear to be the case with the i7-4770's L1.  

Such simple caches -- particularly without the index hashing mentioned above -- can be 
vulnerable to excessive index conflicts, but \malloc{} allocators can be 
made \emph{index-aware}~\cite{ISMM11-Afek-CIA} to mitigate and reduce the 
frequency of index conflicts. Index imbalance results in underutilization of 
the cache. Some indices will be ``cold'' (less frequently accessed) while others 
are ``hot'' and oversubscribed and thus incur relatively higher miss rates. It's worth pointing out 
that most application/allocator combinations don't exhibit excessive index 
conflicts, but for those that do, the performance impact can be significant. 
An index-aware allocator can act to ``immunize'' an application against some 
common cases of index-imbalance while typically incurring no additional cost 
over index-oblivious allocators. 
\citet{ISMM11-Afek-CIA} describes an index-aware allocator designed for the 
L1 in a SPARC T2+ processor, but the changes required to retarget 
retarget the allocator to the cache geometry of the i7-4770 are trivial.
\Invisible{Cache geometry} 
\Invisible{Think of index-aware allocator as cheap insurance against a
rare but painful performance disorder}  

The \emph{CIA-Malloc} (Cache-Index Aware) allocator described in ~\cite{ISMM11-Afek-CIA} 
has a number of other useful design properties. 
It also happens to be NUMA-friendly and large-page-friendly. Underlying pages are 
allocated on the node where the \malloc{} operation was invoked. Put another way, 
the pages underlying a block returned by \malloc{} will typically reside on the 
node where the \malloc{} was invoked. The allocator is also scalable with very 
little internal lock contention or coherence traffic. Each per-CPU sub-heap 
has a private lock -- the only source of contention is via migration 
or preemption, which are relatively rare. The critical sections are also 
constant-time and very short.  The implementation also makes heavy use of the \texttt{trylock} 
primitive, so if a thread is obstructed it can usually make progress by 
reverting to another data structure. Remote \texttt{free} operations are lock-free. 
In additional to acting to distribute blocks over cache indices -- reducing index
imbalance -- the allocator also tends to more equitably distribute allocated blocks 
over coherence planes~\cite{planes}, cache banks and DRAM channels, resulting in 
reduced channel congestion.  
Critically, the allocator acts to reduce the cost of \malloc{} and \texttt{free} 
operations as well as the cost to the application when accessing blocks allocated 
via \malloc{}. The allocator is also designed specifically to reduce common cases 
of false sharing : allocator metadata-vs-metadata; metadata-vs-block; and 
inter-block block-vs-block. Metadata-vs-metadata sharing and false sharing is 
reduced by using per-CPU sub-heaps. False sharing arising between adjacent 
data blocks -- blocks returned by \malloc{} -- is addressed by placement and alignment. 
These attributes will prove even more useful when we use CIA-Malloc in conjunction 
with hardware transactions.  Specifically, allocator-induced false sharing results in so-called
\emph{coherence misses} in normal execution mode, but in transactional mode those
misses translate into aborts, which are typically more expensive than cache misses.  

The i7-4770 provides hardware transactional memory (HTM), the implementation
of which is similar to that in Sun's \emph{ROCK} processor~\cite{asplos09-dice}.
Our particular interest
is in the use of \emph{Restricted Transactional Memory} (RTM) for the purposes of
TLE.  The critical section body contains unmodified HTM-oblivious legacy code 
that expects to run under the lock in the usual fashion, but via TLE we can 
modify the lock implementation to attempt optimistic execution, reverting to classic
\emph{physical locking} only as necessary. The i7-4770's HTM implementation tracks the 
transactional write-set in the L1 and the read-set in the L3 . It uses a 
requester-wins conflict resolution strategy implemented via the coherence protocol. 
At most a single cache can have a given line in \emph{modified} or \emph{exclusive} 
state at any one time -- a classic multiple-reader single-writer model. Eviction 
or invalidation of a tracked cache line results in a transactional abort.  
For example if a transaction on CPU $C$ loads address $A$, and some other CPU writes 
$A$ before $C$ commits, the write will invalidate the line from $C's$ cache and cause 
an abort. Similarly, if $C$ stores into $A$ and some other CPU loads or stores 
into $A$ before $C$ commits, the invalidation of $A$ will cause $C's$ transaction to 
abort. Read-write or write-write sharing on locations accessed within a 
transaction results in coherence invalidation and consequent abort. 
\footnote{We caution the reader about confusing terminology.  A \emph{data conflict
abort} occurs when a transaction running on CPU $A$ reads a location on some cache 
line $L$ and another CPU $B$ subsequently -- but before $A's$ transaction
can commit -- writes into $L$ \emph{or} if $A$ writes to $L$ in a transaction and $B$ 
concurrently reads or writes to $L$ before $A$ commits.  Put another way,
if CPU $A$ has $L$ in its read or write set, and accesses by CPU $B$ invalidate
$L$ from $A's$ cache, then $A's$ transaction will consequently abort.
Critically, aborts arising from \emph{conflict misses} are distinct 
from \emph{conflict aborts}.} 

In addition to coherence traffic, self-displacement via conflict misses can 
also result in aborts. This is where a CIA-Malloc allocator may provide benefit 
relative to other allocators. Normally an index-aware allocator is expected to 
reduce conflict misses arising from index-imbalance, but it can also reduce 
transactional aborts caused by eviction of read-set or write-set entries from 
index conflicts. Aborts are usually far more expensive than simple cache misses. 
(Absent any potential benefit from warming up of caches, aborts are purely wasted 
and futile effort).

The closest related work is that of
\citet{ppopp15-baldassin}, which explores the impact of \malloc{} allocator implementations
on the performance of applications that use software transactional memory.  They
do not address the interplay between hardware transactional memory and
allocator placement, however. 
\Invisible{Interplay : allocator; STM subsystem; application} 

\section{Evaluation} 

We now show some examples of the influence of virtual address placement
on HTM aborts.  All tests were run on an Intel i7-4770 processor with
turbo mode disabled and sufficient cooling capacity to avoid any thermal throttling. 
The i7-4770 has 4 cores with 2 virtual ``hyperthreads'' per core and runs
at 3.4GHz.  The system was running Ubuntu 14.10 with a Linux 3.16 kernel.  
All applications and libraries were written in C or C++ and compiled with 
gcc 4.9.1 in 64-bit mode.

\begin{figure*}[hbtp]
\begin{center}
\includegraphics[angle=270,origin=c,width=16cm]{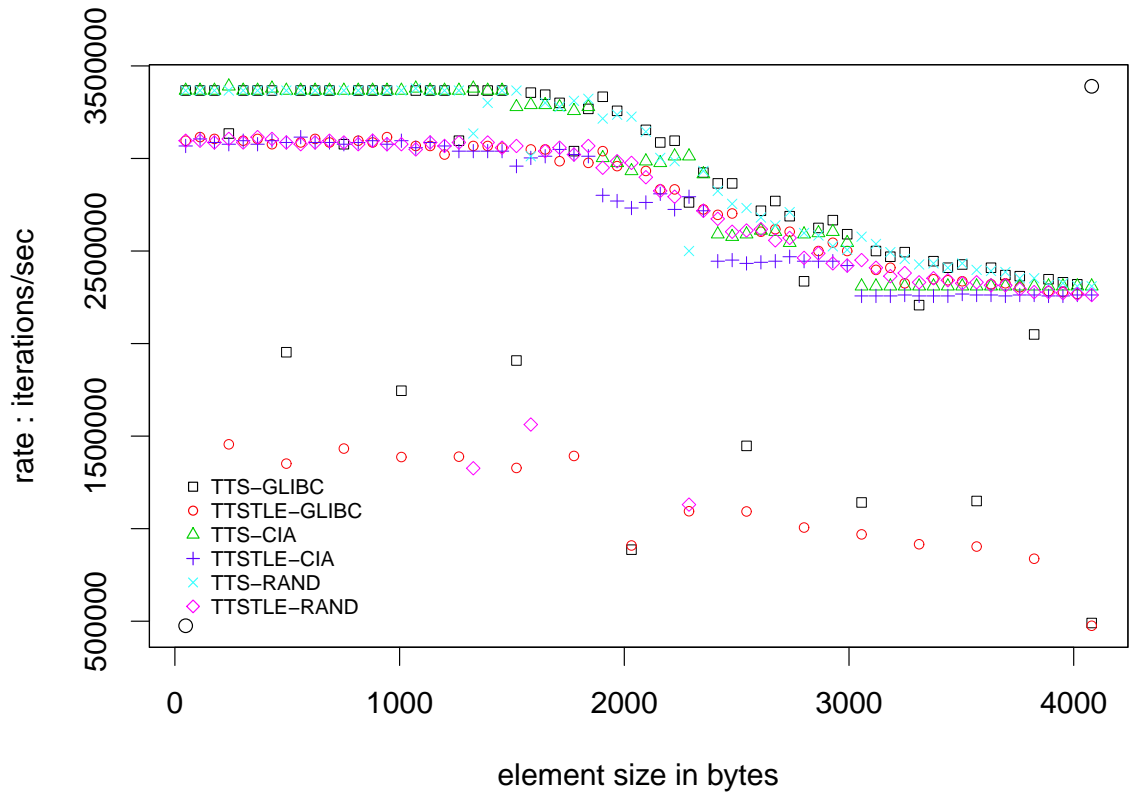}
\end{center}
\caption{Single-threaded ring traversal rates} 
\label{figure:ringer}
\end{figure*}

In Figure \ref{figure:ringer} we use a single-threaded microbenchmark which, 
at startup, allocates a set of 128 nodes.  
Each node has a \emph{Next} field at 
offset 0 followed by a 32-bit integer field \emph{Value}.  The benchmark allocates each
node individually via \malloc{} and then organizes the nodes into a 
intrusively circularly linked list via the \emph{Next} field.  
Since there is a correlation between allocation order and virtual address,
we randomize the order of the nodes with a Fisher-Yates shuffle in order
to minimize the impact of automatic hardware stride-based prefetchers. 
Such a randomized order can put additional stress on the translation lookaside buffers 
(TLBs) by increasing the number of page crossing in a given traversal of the ring.
However unless otherwise stated, TLB misses are not a dominant influence in our results.  
The benchmark then times 10 million traversals of the ring, where each traversal first calls
\texttt{pthread\_mutex\_lock} to acquire a lock, traverses the list, and then
releases that lock with \texttt{pthread\_mutex\_unlock}.  Each step of the 
enclosed loop body executes the following :
\begin{verbatim}    w = w->Next; w->Value = 0 ; \end{verbatim} 
At the end of the run the microbenchmark reports the iteration rate. 
(In this context, an ``iteration'' refers to the act of acquiring the
lock, traversing the full circumference of ring, and finally releasing the lock). 
Crucially, there there are no allocations or deallocations during the measurement
interval.  Instead, the benchmark times accesses to a set of objects that were previously
allocated via \malloc{}. 
In the graph we report the median of 5 separate runs.  The x-axis is the node size, 
which can be controlled via a command-line argument.  The y-axis reflects the 
traversal rate expressed as iterations per second of the ring.  (The 
microbenchmark also reports additional details such as distribution of node
base addresses over the L1 indices).  Since there is just one thread, the lock is never 
contended and the thread never waits. 

Note that only the \emph{Next} and \emph{Value} fields are
accessed.  The remainder of the element is not accessed during the measurement
interval.  Such access patterns are not 
uncommon and can be found in various lookup structures where headers are iterated
over but the larger body of an object is less likely to be accessed.    

We plot 6 sets of points varying combinations of 3 \malloc{} allocators and 2 lock
implementations.   \textbf{TTS} is an LD\_PRELOAD library that 
interposes on the \texttt{pthread\_mutex} family of operators and implements
a simple test-and-test-and-set spin lock.  \textbf{TTSTLE} is just \textbf{TTS}
augmented with simplistic TLE.  All coherence conflict aborts are retried indefinitely.
Unresolvable aborts -- such as those arising from conflict misses underlying the L1
write set -- revert to the slow path and traditional \textbf{TTS} locking.  To avoid the \emph{lemming effect}
~\cite{SMLI-TR-2009-180} we use unbounded spinning to wait for the lock before
trying or retrying the fast path.  \textbf{GLIBC} is the default GNU \texttt{libc}
\malloc{} allocator. \textbf{CIA} is an implementation of the index-aware
allocator described in ~\cite{ISMM11-Afek-CIA} but has been modified to
use the L1 geometry of the i7-4770 and to 
use size classes that are prime multiples of the cache line size (64 bytes).  
This helps avoid both intra- and inter-size class index conflicts   
\footnote{The original \textbf{CIA} allocator used so-called \emph{punctuated arrays} but
for these experiments our implementation avoided that technique and instead
depends on the prime-based size class policy noted above.}.   
\textbf{CIA} is implemented as an LD\_PRELOAD interposition library.   
Finally, \textbf{RAND} is an interposition library that intercepts \malloc{} calls and
probabilistically adds a small number to the requested size, and then passes 
control to the underlying \malloc{} in \textbf{GLIBC}.  
Such randomization can intentionally introduce irregularity
into the spacing of blocks and act to reduce index conflicts.   We include 
\textbf{RAND} because it provides some degree of relief with rather trivial
overheads and an extremely simple implementation. 
\Invisible{palliative; remedial; provides relief} 
\Invisible{probabilistic; Bernoulli trial} 
\Invisible{causality analysis; etiology} 
\Invisible{Confounding factor} 
\Invisible{Tesselation; tiling; stride} 

As can be seen in Figure \ref{figure:ringer} we can find a subset of points where 
\textbf{TTS-GLIBC} significantly underperforms \textbf{TTS-CIA} and the main sequence.
Degraded performance occurs near element sizes of 512, 1K, 1.5K, 2K, 2.5K, 3K and 
4K bytes, for instance.  Using hardware performance counters we find that the
degraded performance correlates with increased L1 miss rates, supporting our
claim that those sizes are index-unfriendly and result in index imbalance and
underutilization of L1.  When the stride between nodes is 1K, for instance,
node base addresses map to just 4 of the possible 64 L1 indices, resulting
in potential imbalance and under-utilization of the L1.  In more detail,
say we have a collection of $N$ elements, each of which was allocated via 
\texttt{malloc($S$)}. The allocator may place those $N$ objects in a contiguous
fashion such that the values returned by \texttt{malloc($S$)} differ by $\bar{S}$.
$\bar{S}$ may be greater than $S$ because of quantization and potential per-block
\texttt{malloc} metadata headers and footers.   If $\bar{S}$ -- the effective stride -- 
happens to be 1K, for instance, then the $N$ blocks may fall on just 4 of the 
possible 64 L1 indices, resulting in conflict misses as we traverse the collection. 

We also observe that \textbf{TTSTLE-GLIBC} underperforms
\textbf{TTS-GLIBC} at those same points.  Under \textbf{TTSTLE}, conflict
misses cause the fast path transaction to abort with an ``internal buffer overflow'' 
error code \footnote{To the best of our knowledge, this abort code indicate self-displacement
of read-set or write-set elements}.  
This abort is not generally retryable, so \textbf{TTSTLE} reverts to the 
non-transactional normal mode slow path.  
The transactional attempt was futile and constituted wasted effort.  In particular, 
unresolvable aborts are more costly than cache misses.  Generally,
the \textbf{CIA} forms outperform \textbf{RAND} which in turns outperforms \textbf{GLIBC}.

The inflection point in the main sequence at about 2000 bytes arises from level-1 data
TLB misses.  For the default 4KB page size, the i7-4770 has a 64-entry 4-way set associative
level-1 data TLB (L1-DTLB) and a 1024-entry 8-way set associative level-2 
unifield TLB (L2-TLB).  
The L1-DTLB thus has a maximum ``span'' of 256KB (64 TLB entries $*$ 4KB pages).  
With 128 elements of 2000 bytes each, the best-case minimum TLB footprint of 
the ring is 256KB, matching the capacity of the L1-DTLB.  
All the allocators provide reasonably dense and compact placement of the ring elements.  
The cache footprint of the ring -- the number of lines underlying the \emph{Next}
and \emph{Value} fields -- depends in part on the alignment of blocks returned
by \malloc{}.  \textbf{CIA} always returns addresses aligned on 64-byte boundaries
while the default \textbf{GLIBC} allocator -- and consequently the \textbf{RAND} 
allocator -- return addresses only guaranteed 8-byte alignment.  Under \textbf{CIA}
both the \emph{Next} and \emph{Value} fields reside on the same cache 
line, and the cache footprint is simply the number of elements in the ring multiplied
by the cache line size of 64-bytes.  That is, the cache footprint of  
the ring is independent of the element size.  With 128 elements, the cache 
footprint is just 8KB, or $1/4$ of
the L1's capacity.  With worst-case pessimal alignment, under \textbf{RAND} and
\textbf{GLIBC} the \emph{Next} and \emph{Value} fields will be split and reside
on two adjacent lines.   In that case the cache footprint would be 16KB or $1/2$
the L1's capacity.  If the L1 were an ideal fully associative cache, the ring would
fit comfortably in the L1, and subsequent traversal could be completed with any misses.
But because of index conflicts, traversals may be subject to conflict misses. 

Broadly, the \textbf{GLIBC} allocator exhibits reduced performance at 
certain pathological sizes.   At such problematic sizes, \textbf{TTSTLE} underperforms
\textbf{TTS} by a significant margin because of cycles wasted on futile
transactions.  \textbf{RAND} provides some benefit relative to \textbf{GLIBC}
but in some cases yields poor performance.  \textbf{CIA} avoids the pathological
sizes completely.   We note in passing that two horizontal ``bands'' appear 
in the figure on the left-hand side of the graph.  Transactional executing appears
to be slightly slower than normal execution.   We believe this is an artifact
of higher latencies associated with TSX than occur with the normal atomic
operations used to acquire and release a mutex. 

\Invisible{HR Diagram; Hertzsprung–Russell; main sequence} 
\Invisible{Ring circumference}  
\Invisible{Allocators that use sizes classes result in size quantization and 
some internal fragmentation ; wastage} 
\Invisible{cyclic access; scan resistance; pseudo-LRU; 
cache replacement policy;
Intel L3 is not pure-strict LRU 
}

\begin{figure*}[hbtp]
\begin{center}
\includegraphics[angle=270,origin=c,width=16cm]{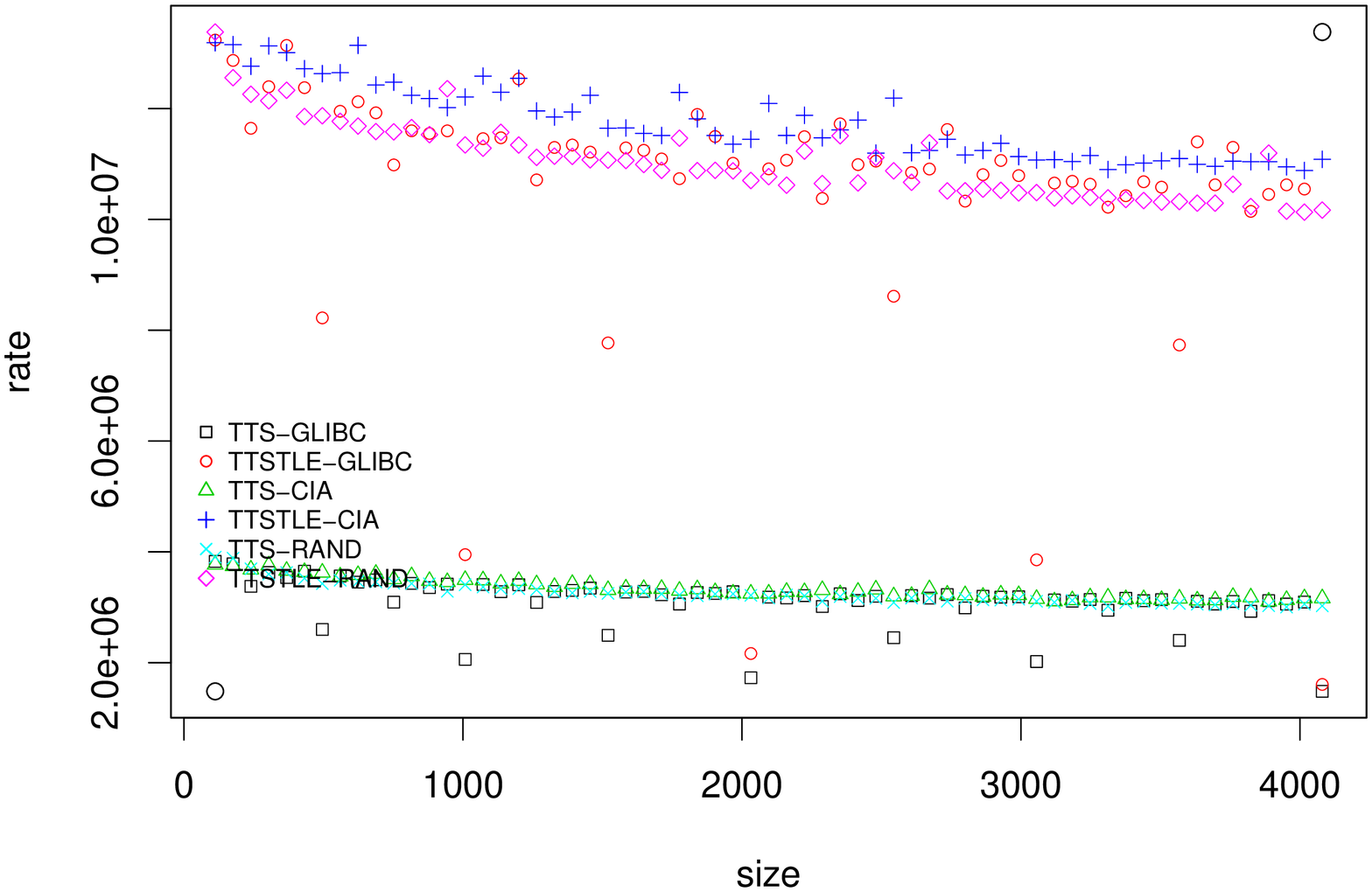}
\end{center}
\caption{AVL tree throughput with 4 threads} 
\label{figure:avl} 
\end{figure*}

In Figure \ref{figure:avl} we show how the problem of aborts arising from
conflict misses is amplified when using TLE with multiple concurrent threads.  In particular,
conflict misses cause aborts and aborts force the lock to use the classic slow path, 
greatly reducing the opportunities for concurrency.  

For the benchmark presented in
Figure \ref{figure:avl} we use a single shared AVL tree ~\cite{Cormen, AVLTree} where 
insert, delete and update operations are protected by a single 
\texttt{pthread\_mutex} instance.  The AVL tree is based on the implementation
used in OpenSolaris ~\cite{AVLSolaris}.  The only changes to the AVL tree code iself
were to (a) insert padding (alignment constraints) between the frequently updated element 
count field and other fields in the tree descriptor structure; and (b) to move the update
of that element count to the end of the critical section.  These were the only
concessions to make the AVL tree code more transaction-friendly.
Sequestering the count field as the sole occupant of its own cache sector acts to reduce false sharing 
and consequent transactional aborts \footnote{Because of the \emph{adjacent sector prefetch}
facility, we align to 128 bytes instead of 64 bytes, even though 64 bytes is the line
size throughout the cache hierarchy and the unit of coherence.  The Intel manuals also recommend
128 bytes for the purposes of avoiding false sharing}. 
In addition, new structural and content integrity check routines were added. 
These are used after a run to check the validity of the tree. 
The tree is intrusively linked and each node is individually allocated 
and freed via \texttt{malloc} and \texttt{free} calls.  These allocation 
and deallocation operations are executed within the critical section.  
The AVL tree implements a key-value store, with the key and value both 
being 32-bit integers. Each AVL tree node contains AVL tree linkage,
a key field, a value field, and a variable size area.  (That area is never
accessed).  The size of the AVL tree node
is controlled by a command-inline argument and is reflected in the x-axis. 
The benchmark spawns 4 concurrent threads, each of which loops by generating
random numbers -- via a thread-local uniform pseudo-random number generator -- 
to control the operation type and key \footnote{We opted to use 4 threads to
avoid hyperthreaded execution --  with 4 threads we have just 1 thread per core} :
30\% of the time the loop will insert a new key (if the key
already exists in the tree, its value is updated); 30\% of the time a key is deleted
and 40\% of the time a lookup is performed.  The key range is $[0-65536)$.  The
tree is initially populated to half capacity (32767 elements) with a random set of keys.  At the 
end of a 10 second measurement interval the benchmark reports the aggregate
operation rate.  (An operation is an iteration of the loop that inserts, deletes
or looks up keys in the tree).  This rate is shown on the y-axis and expressed
in operations per second.   We report the median of 5 independent runs.  

Not surprisingly, the \textbf{TTSTLE} forms outperform the \textbf{TTS} forms, 
as more concurrency is available.  But again, for \textbf{TTSTLE-GLIBC} we
see the same set of pathological sizes as was found in Figure \ref{figure:ringer}. 
At about 2000 bytes, for instance, \textbf{TTSTLE-GLIBC} with 4 threads actually
performs worse than than the best of the serialized \textbf{TTS} forms.  
This reflects the compounding effect of restricted concurrency and wasted cycles
in futile transactions that end in abort. 


\section{Conclusion}

We have shown that the use of index-aware allocators can avoid certain
pathological cases where index conflicts cause misses, aborts, and
potentially restrict concurrency under TLE.   

Put simply, allocator placement can influence conflict miss rates, which
in turn influence abort rates, which in turn can force threads to abandon
fast path TLE execution and revert to serializated execution under a lock,
restricting parallelism.  An index-aware allocator can provide some relief
against this phenomena.  Absent such an allocator, randomization of sizes
at either the allocation size or in the allocator (\textbf{RAND}) may provide 
benefit by distrupting regularity in placement.  

Programming with hardware transactional memory is in it infancy, so the 
degree to which programs might be afflicted by aborts arising from
index conflicts is unknown.  Generally, we expect the problem to be infrequent,
but when it does manifest, the impact can be surprising and significant.  We 
suggest index-aware allocators as a way to reduce the odds of encountering
the problem.  

Hardware-based remedies to reduce the rate of conflict misses were suggested          
Seznec \cite{SkewAssociative} (skew-associative caches) and later by                 
by Gonzales \cite{263599} and Wang \cite{NewCache} and Sanchez \cite{ZCache}. 
All require changes to the hash function that maps addresses to cache indices. 
By acting to reduce conflict misses, they would also reduce aborts arising
from such misses. 

We note in passing that under a requester-wins conflict resolution strategy --
as if found with the current members of the ``Haswell'' family -- to the extent 
possible and reasonable it is useful to shift stores of frequently 
accessed shared variables toward the end of a transaction ~\cite{Shifting}.  
This might be accomplished by hand, or a transaction-aware compiler or just-in-time
compiler (JIT) can perform 
some of the transformations.  Shifting reduces the window of vulnerability 
where the store resides in the transaction's write-set. (Active transactions
are vulnerable in both time and space).  But the asymmetry in 
the i7-4770 where the write-set is tracked in the L1 and the read-set in the 
L1, L2 and L3 gives us yet another reason to shift stores toward the end of a 
transaction. Consider a transaction that executes a store followed by large 
number of loads. Those loads may displace the store from the L1 and cause an abort. But if 
we shift the store to the end of the transaction, the same set of accesses 
(just reordered) can succeed without abort. The store may displace a loaded 
line from the L1, but the L2 and L3 can still track the line.






\bibliographystyle{abbrvnat} 
\bibliography{main}

\begin{thebibliography}{20}
\providecommand{\natexlab}[1]{#1}
\providecommand{\url}[1]{\texttt{#1}}
\expandafter\ifx\csname urlstyle\endcsname\relax
  \providecommand{\doi}[1]{doi: #1}\else
  \providecommand{\doi}{doi: \begingroup \urlstyle{rm}\Url}\fi

\bibitem[AVL()]{AVLSolaris}
{Solaris AVL Tree Implementation}.
\newblock URL
  \url{https://github.com/illumos/illumos-gate/blob/master/usr/src/uts/common/sys/avl_impl.h}.

\bibitem[Afek et~al.(2011)Afek, Dice, and Morrison]{ISMM11-Afek-CIA}
Y.~Afek, D.~Dice, and A.~Morrison.
\newblock Cache index-aware memory allocation.
\newblock In \emph{Proceedings of the International Symposium on Memory
  Management}, ISMM '11, 2011.
\newblock URL \url{http://doi.acm.org/10.1145/1993478.1993486}.

\bibitem[Baldassin et~al.(2015)Baldassin, Borin, and Araujo]{ppopp15-baldassin}
A.~Baldassin, E.~Borin, and G.~Araujo.
\newblock Performance implications of dynamic memory allocators on
  transactional memory systems.
\newblock In \emph{Proceedings of the 20th ACM SIGPLAN Symposium on Principles
  and Practice of Parallel Programming}, PPoPP 2015, 2015.
\newblock URL \url{http://doi.acm.org/10.1145/2688500.2688504}.

\bibitem[Cormen et~al.(2001)Cormen, Stein, Rivest, and Leiserson]{Cormen}
T.~H. Cormen, C.~Stein, R.~L. Rivest, and C.~E. Leiserson.
\newblock \emph{{Introduction to Algorithms}}.
\newblock McGraw-Hill Higher Education, 2nd edition, 2001.
\newblock ISBN 0070131511.

\bibitem[Dice(2008)]{Shifting}
D.~Dice.
\newblock {Reducing Transactional Abort Rates Using Compiler Optimization
  Techniques}, 2008.
\newblock URL \url{http://www.google.com/patents/US20100169870}.
\newblock {US Patent Application -- US20100169870}.

\bibitem[Dice(2015)]{RWTLE}
D.~Dice.
\newblock {Using reader-writer locks to improve hardware TLE}, 2015.
\newblock URL
  \url{{https://blogs.oracle.com/dave/entry/using_reader_writer_locks_to}}.

\bibitem[Dice et~al.(2009{\natexlab{a}})Dice, Lev, Moir, and
  Nussbaum]{asplos09-dice}
D.~Dice, Y.~Lev, M.~Moir, and D.~Nussbaum.
\newblock {Early Experience with a Commercial Hardware Transactional Memory
  Implementation}.
\newblock In \emph{Proceedings of the 14th International Conference on
  Architectural Support for Programming Languages and Operating Systems},
  ASPLOS XIV, 2009{\natexlab{a}}.
\newblock URL \url{http://doi.acm.org/10.1145/1508244.1508263}.

\bibitem[Dice et~al.(2009{\natexlab{b}})Dice, Lev, Moir, Nussbaum, and
  Olszewski]{SMLI-TR-2009-180}
D.~Dice, Y.~Lev, M.~Moir, D.~Nussbaum, and M.~Olszewski.
\newblock {Early Experience with a Commercial Hardware Transactional Memory
  Implementation}, 2009{\natexlab{b}}.
\newblock URL
  \url{https://blogs.oracle.com/dave/resource/smli_tr-2009-180.pdf}.
\newblock Sun Labs Technical Report SMLI TR--2009--180.

\bibitem[Dice et~al.(2014)Dice, Harris, Kogan, Lev, and Moir]{wttm14-dice}
D.~Dice, T.~L. Harris, A.~Kogan, Y.~Lev, and M.~Moir.
\newblock {Pitfalls of Lazy Subscription}, 2014.
\newblock URL
  \url{https://blogs.oracle.com/dave/resource/wttm14-dice-PitfallsLazySubscription.pdf}.
\newblock 6th Workshop on the Theory of Transactional Memory -- EuroTM WTTM
  2014.

\bibitem[Gonz\'{a}lez et~al.(1997)Gonz\'{a}lez, Valero, Topham, and
  Parcerisa]{263599}
A.~Gonz\'{a}lez, M.~Valero, N.~Topham, and J.~M. Parcerisa.
\newblock Eliminating cache conflict misses through xor-based placement
  functions.
\newblock In \emph{ICS '97: Proceedings of the 11th international conference on
  Supercomputing}, 1997.
\newblock URL \url{http://doi.acm.org/10.1145/263580.263599}.

\bibitem[Hetherington and Phillips(2008)]{planes}
R.~Hetherington and S.~Phillips.
\newblock Multiple independent coherence planes for maintaining coherency,
  2008.
\newblock URL \url{http://www.google.com/patents/US7353340}.
\newblock US Patent 7,353,340.

\bibitem[Hill and Smith(1989)]{Hill89}
M.~D. Hill and A.~J. Smith.
\newblock {Evaluating Associativity in CPU Caches}.
\newblock \emph{IEEE Trans. Comput.}, 38\penalty0 (12), Dec. 1989.
\newblock URL \url{http://dx.doi.org/10.1109/12.40842}.

\bibitem[Hund et~al.(2013)Hund, Willems, and Holz]{Hund}
R.~Hund, C.~Willems, and T.~Holz.
\newblock Practical timing side channel attacks against kernel space aslr.
\newblock In \emph{Security and Privacy (SP), 2013 IEEE Symposium on}, May
  2013.
\newblock URL \url{http://dx.doi.org/10.1109/SP.2013.23}.

\bibitem[{Intel Corporation}(2012)]{Haswell}
{Intel Corporation}.
\newblock {Transactional Synchronization in Haswell}, 2012.
\newblock URL
  \url{https://software.intel.com/en-us/blogs/2012/02/07/transactional-synchronization-in-haswell/}.
\newblock [online; retrieved 2015].

\bibitem[Ritson and Barnes(2013)]{RitsonBarns}
C.~G. Ritson and F.~R. Barnes.
\newblock {An evaluation of Intel's Restricted Transactional Memory for CPAs},
  2013.
\newblock URL
  \url{http://www.wotug.org/papers/CPA-2013/RitsonBarnes13/RitsonBarnes13.pdf}.

\bibitem[Romer et~al.(1994)Romer, Lee, Bershad, and Chen]{Romer94}
T.~Romer, D.~Lee, B.~N. Bershad, and J.~B. Chen.
\newblock {Dynamic Page Mapping Policies for Cache Conflict Resolution on
  Standard Hardware}.
\newblock In \emph{In 1st USENIX Symposium on Operating Systems Design and
  Implementation (OSDI)}, pages 255--266, 1994.

\bibitem[Sanchez and Christos(2010)]{ZCache}
D.~Sanchez and K.~Christos.
\newblock The zcache: Decoupling ways and associativity.
\newblock In \emph{MICRO 43: Proceedings of the 43rd Annual IEEE/ACM
  International Symposium on Microarchitecture}. IEEE Computer Society, 2010.

\bibitem[Seznec(1993)]{SkewAssociative}
A.~Seznec.
\newblock A case for two-way skewed-associative caches.
\newblock In \emph{ISCA '93: Proceedings of the 20th annual international
  symposium on Computer architecture}, 1993.
\newblock URL \url{http://doi.acm.org/10.1145/165123.165152}.

\bibitem[Wikipedia()]{AVLTree}
Wikipedia.
\newblock {AVL Tree}.
\newblock URL \url{http://en.wikipedia.org/wiki/AVL_tree}.

\bibitem[Zhenghong and Lee(2008)]{NewCache}
W.~Zhenghong and R.~B. Lee.
\newblock A novel cache architecture with enhanced performance and security.
\newblock In \emph{MICRO 41: Proceedings of the 41st annual IEEE/ACM
  International Symposium on Microarchitecture}, 2008.
\newblock URL \url{http://dx.doi.org/10.1109/MICRO.2008.4771781}.

\end{thebibliography}







\end{document}